# Automatic Identification of Subjects for Textual Documents in Digital Libraries

*Kuang-hua Chen*

Department of Library and Information Science
National Taiwan University
1, SEC. 4, Roosevelt RD., Taipei
TAIWAN, 10617, R.O.C.
E-mail: khchen@ccms.ntu.edu.tw

**ABSTRACT**
The amount of electronic documents in the Internet grows very quickly. How to effectively identify subjects for documents becomes an important issue. In past, the researches focus on the behavior of nouns in documents. Although subjects are composed of nouns, the constituents that determine which nouns are subjects are not only nouns. Based on the assumption that texts are well-organized and event-driven, nouns and verbs together contribute the process of subject identification. This paper considers four factors: 1) word importance, 2) word frequency, 3) word co-occurrence, and 4) word distance and proposes a model to identify subjects for textual documents. The preliminary experiments show that the performance of the proposed model is close to that of human beings.

**INTRODUCTION**
Due to the increasingly growth of Internet, the amount of electronic resources accumulates very quickly. In addition, the commercial systems go into the Internet and show off themselves as soon as possible. All of these phenomena contribute the development of Internet society and make us have a virtual image of a global village. As a result, the transfer of information becomes fast and human beings enjoy great information services, which are never seen before. In past, the information providers were libraries, museums, and organizations. Nowadays, everybody can be an information provider that publishes his/her essays, pieces of works, papers, or the like on the Internet.

According to some statistics, the amount of information doubles every 20 months. Note that this was estimated in 1991. The traffic of Usenet doubles each year and that of information in Internet increases 12% each month [15]. As to Taiwan, the growth rate of the backbone of academic network TANet is 1000 percent from 1992 to 1995. Now, users are confronted with the dilemma of finding useful information rather than the lack of information.

In order to help users find useful information or documents from immense datahouses like Internet, some kind of value-added processing is indispensable. Language techniques are widely used in researches of information retrieval (IR) and make great improvements in the performance of IR systems. Although Internet documents usually appear in multimedia form, the majority of documents are the collection of written languages. Therefore, it makes sense to apply language techniques to analyze Internet documents.

This paper focuses on identifying subjects for textual documents. The subjects of documents, which are automatically identified by the proposed model, could be used to describe documents and to be one attribute of metadata. As a result, the proposed model could help Internet users find what they want in an effective way. Four factors are considered in constructing the model: word importance, word frequency, word co-occurrence, and word distance. In addition, nouns and verbs take part in the proposed model rather than nouns only in other models.

The rest of this paper is structured as follows. Section 2 discusses the relative works. Section 3 presents a corpus-based model and describes the training process. Section 4 shows a series of experiments and analyzes the experimental results. Section 5 is a short conclusion.

**BACKGROUND AND RELATED WORKS**
The collections of libraries are always value-addedly processed to help readers or users access the needed information. For example, indexers and abstractors associate index terms and abstracts to each document-like object (DLO). In general, each type of DLO has its own appropriate metadata to describe various attributes for different purposes. The MARC (Machine-Readable Cataloging) Format [9] used in libraries could be regarded as a kind of metadata formats, which records very complicated data. In contrast, Dublin Core [20] consists of only fifteen necessary attributes for describing Internet recourses. No matter what kind of metadata is used, there is usually one attribute for recording subjects of a DLO. The Library of Congress uses Library of Congress Subject Headings [12] as the controlled vocabularies for subject cataloging. The National Central Library in Taiwan publishes Chinese Subject Headings [5] for the same purpose.

The work of subject assignment is a mental activity. Librarians conform to the policies of their libraries and choose some subjects from Subject Headings. This effort takes many human costs and is a time-consuming work. The use of language techniques will effectively reduce the costs and quickly fulfill the tasks.

In tradition, term frequency (TF) is widely used in researches of information retrieval. The idea is that after excluding the functional words, the words occurring frequently would carry the meaning underlying a text. However, if these words appear in many documents, the discriminative power of words will decrease. Sparck Jones [19] proposed inverse document frequency (IDF) to rectify the aforementioned shortcoming. The IDF is shown as follows:

$$IDF(w) = \log(P-O(w))/O(w),$$

where $P$ is the number of documents in a collection, $O(w)$ is the number of documents with word $w$. The combination of TF and IDF denotes a great success in the information retrieval and effectively promotes the performance of IR systems.

Salton and his colleagues [16-18] extended the concept of IDF and devised term discrimination value and other parameters. Their works also improved IR systems.

Some researches of text modeling have been reported in literature. Grosz and Sidner proposed rhetorical discourse structure to model texts [6]. They focused on the impacts of attention and intention on the discourse structure. Morris and Hirst proposed five types of thesaural relations to find lexical chains and use these chains to determine the structure of text [14]. Some demerits remain. First, the strength of thesaural relation of a word pair is not the same as one another. Just to find the relation of two words is not enough. A measure of the strength of thesaural relation should be invented. Second, the algorithm to finding the structure of text is performed manually rather than automatically.

Little literature touches on how to identify subjects automatically. Chen proposed a corpus-based model to deal with the tasks. However, the computation time is large and is not practical for Internet applications [1]. Lin used a knowledge-based approach to count the central ideas in a text. The shortcoming is that the approach depends on a well-formed hierarchical concept taxonomy and this taxonomy does not exist for many languages [13]. No matter what kinds of approaches are used, the subject identification is important and practical for many researches, e.g., anaphora resolution, information retrieval, text summarization, and so on. The identified topics could form the *universe* in DRT [11]. Therefore, the antecedents of anaphora could be chosen from the universe. For information retrieval, a few researchers have found that subtopic structures are very useful in retrieving relevant documents [7]. Identifying subjects could be the first step in the automatic summarization [2].

**THE PROPOSED MODEL**

Nouns and verbs in a well-organized text are coherent in general. We postulate that

*Noun-verb is a predicate-argument relation-ship on sentence level and noun-noun relationship is associated on discourse level.*

The following text is an example.

*I want to arrange a meeting with Ira. It should be held at 3 p.m. We can get together in his office. Invite John to come, too.*

Nouns and verbs in this text are related to one another. The aforementioned relationships may be represented implicitly via collocational semantics. In order to identify the subjects of a document, the first step is to analyze the factors of composing texts. That is, the writing process of human beings. Based on the postulation mentioned above, we use four distributional parameters to construct a text model:

- Word importance
- Word frequency
- Word co-occurrence
- Word distance

The following will discuss each factor in sequence.

The word importance means that when this word appears in texts, how strong it is to be the core word of texts. In other words, it represents the possibility of selecting this word as an index term. The IDF is chosen to measure the word importance in this paper. The bigger the collection is, the more stable the IDF is.

In addition, the frequency of a word itself does also play an important role in texts. For example, the word with high frequency usually makes readers impressive. The proposed model combines the two factors as the predecessors did.

If a text discusses a special subject, there should be many coherent words together to support this subject. That is to say, these words will co-occur frequently. From the viewpoint of statistics, some kind of distributional parameters like mutual information [3] could be used to capture this phenomenon.

Including the distance factor is motivated by the fact that related events are usually located in the same texthood. This is the spatial locality of events in a discourse. The distance is measured by the difference between cardinal numbers of two words. We assign a cardinal number to each verb and noun in sentences. The cardinal numbers are kept continuous across sentences in the same paragraph. The following is an example.

With so many problems$_1$ to solve$_2$, it would be a great help$_3$ to select$_4$ some one problem$_5$ which might be the key$_6$ to all the others, and begin$_7$ there. If there is any such key-problem$_8$, then it is undoubtedly the problem$_9$ of the unity$_{10}$ of the Gospel$_{11}$. There are three views$_{12}$ of the Fourth Gospel$_{13}$ which have been held$_{14}$.

As a result, the distance between two words, $w_1$ and $w_2$, is calculated using the following equation.

$$D(w_1, w_2) = abs(C(w_1) - C(w_2)),$$

where the $D$ denotes the distance and $C$ the cardinal number.

Consider the four factors together, the proposed model is shown as follows:

$$SS(n) = PN \times SNN(n) + PV \times SNV(n)$$

*SS* denotes the score for subject, that is the score for a noun *n* to be a subject, where *SNN* denotes the strength of a noun with other nouns, *SNV* the strength of a noun with other verbs, and *PN* and *PV* are the weights for *SNN* and *SNV*, respectively. The determination of *PN* and *PV* is via deleted interpolation [10] shown as follows.

$$S_V = \sum_i \frac{PV \times SNV(N_i)}{PN \times SNN(N_i) + PV \times SNV(N_i)}$$

$$S_N = \sum_i \frac{PN \times SNN(N_i)}{PN \times SNN(N_i) + PV \times SNV(N_i)}$$

$$PN = \frac{S_N}{S_N + S_V} \qquad PV = \frac{S_V}{S_N + S_V}$$

The equations for *SNV* and *SNN* are shown as follows.

$$SNV(n_i) = \sum_j \frac{IDF(n_i) \times IDF(v_j) \times f(n_i, v_j)}{f(n_i) \times f(v_j) \times D(n_i, v_j)}$$

$$SNN(n_i) = \sum_j \frac{IDF(n_i) \times IDF(n_j) \times f(n_i, n_j)}{f(n_i) \times f(n_j) \times D(n_i, n_j)}$$

$f(w_i, w_j)$ is the co-occurrence of words $w_i$ and $w_j$, and $f(w)$ is the frequency of word $w$. In fact, $f(w_i, w_j)/f(w_i) \times f(w_j)$ (hereafter, mutual frequency, MF in short) is a normalized co-occurrence measure with the same form as the mutual information. Thus, the SNV and SNN could be rewritten as follows:

$$SNV(n_i) = \sum_j \frac{IDF(n_i) \times IDF(v_j) \times MF(n_i, v_j)}{D(n_i, v_j)}$$

$$SNN(n_i) = \sum_j \frac{IDF(n_i) \times IDF(n_j) \times MF(n_i, n_j)}{D(n_i, n_j)}$$

**EXPERIMENTS AND ANALYSES**

The training and the testing corpus are Chinese texts. In order to acquire high-quality Chinese electronic texts as soon as possible, this paper uses the Academia Sinica Balanced Corpus (Sinica Corpus) constructed by CKIP Group [4] as the training corpus. More importantly, the texts of Sinica Corpus are comparable to the Internet documents. Another reason to use Sinica Corpus is that the Chinese sentences which have no word boundary marker should be segmented first before applying any word-based approach. In order to avoid the errors introduced by word segmentation, the word-segmented Sinica Corpus is used to collect the statistical data. Sinica Corpus is the first balanced Chinese corpus. This corpus consists of 5 millions of Chinese words and is carefully organized, labeled with various metadata, and tagged with parts of speech. Each text in the corpus is classified and marked according to five criteria: genre, style, mode, topic, and source. The feature values of these classifications are assigned in a hierarchy.

The experimental procedure consists of training stage, testing stage, and evaluating stage described as follows.

- Training Stage
  1. Calculate IDF values for nouns and verbs
  2. Calculate MF value for each pair of words
  3. Calculate distance for each pair of words
- Testing Stage
  1. Randomly select 10 texts from Sinica Corpus as the testing texts
  2. Use the proposed model to identify ordered subjects for each testing text

- Evaluating Stage
  1. Invite 8 readers to read each text and ask them to choose subjects with no restriction
  2. Compare the subjects selected by readers
  3. Compare the subjects identified by the model and those selected by readers

The statistics of testing texts are shown in Table 1 and one text is shown in Appendix. The format of source tag consists of a file name, a number of the starting sentence, and a number of the ending sentence. For example, _T-SA.TAG:1-13 denotes that this text is extracted from sentence 1 to sentence 13 of the file _T-SA.TAG.

The experimental results are shown in Table 2. Because the model will arrange all nouns of a text in order, Table 2 does not list all nouns but the first 1/3 of nouns. Table 3 shows subjects selected by Reader 4 and Reader 5. Comparing the subjects selected by readers, we find that the phenomenon of inter-indexer inconsistency [8] is obvious in our experiments. Table 4 records mean and standard deviation of the number of chosen subjects among readers. AVG denotes the mean of the number of subjects, and STDEV denotes the standard deviation. Before the readers read texts, the instruction is "please read the texts and assign each text some ordered subjects with no restrictions on number or type of subjects, according to your personal viewpoint." From the Table 4, we observe that the largest STDEV value is 4.01 and the smallest is 0.41 from the viewpoint of readers. However, from the viewpoint of texts, the largest STDEV value is 5.63 and the smallest is 1.36. Although, the effect of inter-indexer inconsistency exists, it is less than that of inter-text inconsistency. Further investigating the repetition of selected subjects, we find that only 7 subjects are selected by all of the 8 readers and 57 subjects are selected by one of the 8 readers. The details are shown in Table 5. This also shows the inconsistent phenomenon among readers.

**Table 1: Testing Texts**

|        | Source Tag         | # of Sentences | # of Words |
|--------|--------------------|----------------|------------|
| TEXT01 | _T-SA.TAG:1-13     | 13             | 102        |
| TEXT02 | _T-SA.TAG:32-50    | 19             | 230        |
| TEXT03 | _T-SY.TAG:1-12     | 12             | 132        |
| TEXT04 | _T-SY.TAG:89-112   | 24             | 222        |
| TEXT05 | _T-SY.TAG:212-244  | 33             | 374        |
| TEXT06 | _T-SL.TAG:60-88    | 29             | 345        |
| TEXT07 | _T-SL.TAG:490-531  | 42             | 406        |
| TEXT08 | _T-SL.TAG:680-705  | 26             | 453        |
| TEXT09 | _T-SL.TAG:721-745  | 25             | 268        |
| TEXT10 | _T-SL.TAG:797-829  | 33             | 353        |

**Table 2: Experimental Results**

|        | The Ordered Subjects Selected by the Model |
|--------|---------------------------------------------|
| TEXT01 | 考試、優待、條文、標準 |
| TEXT02 | 考試、優待、草案、淪陷區、學歷、部長、標準、辦法 |
| TEXT03 | 制度、背後、思想、體系、行為 |
| TEXT04 | 憲法、普莉揚卡、被選舉權、第三世界、復仇者、身分、政權、後起之秀、桑妮雅、時間 |
| TEXT05 | 變化、馬列主義、群眾、死路一條、馬克思、教廷、共產黨人、祖國、口號、卡斯楚 |
| TEXT06 | 史氏、配樂家、ＨＥＲＢＥＲＴ ＳＴＯＴＨＡＲＴ、色彩、民謠、風味、音樂、包機 |
| TEXT07 | 考試、虫子、肢腳、老頭、夢魘、金杏枝、磚頭書、查泰萊、習慣 |
| TEXT08 | 人間、烽火、Ｊａｓｐｅｒｓ、時代、特徵、實質、診斷、商業 |
| TEXT09 | 編輯、點子、朋友、觀光客、過客、市民、刊、感情、城市 |
| TEXT10 | 改革、轉機、政變、軍方、葉爾辛、短文、改革派、機率、保守派 |

**Table 3: The Subjects Selected by Some Readers**

|  | Read 4 | Read 5 |
|---|---|---|
| TEXT01 | 入學考試、優待 | 入學考試、優待、標準 |
| TEXT02 | 學歷、條件、大陸、教育局 | 學歷、學籍、辦法、草案 |
| TEXT03 | 開放、改革、思想、行為、模式、矛盾、轉型期、特權、既得利益、體系 | 轉型期、制度、體制、利益 |
| TEXT04 | 桑妮雅、普莉揚卡、拉吉夫、復仇者、煩惱 | 普莉揚卡、印度、政治、生活 |
| TEXT05 | 卡斯楚、共產黨人、蘇聯、戈巴契夫、馬列主義、第三世界 | 共產黨、古巴、蘇聯、卡斯楚 |
| TEXT06 | 亂世、佳人、美國、霸權、藝術 | 電影、文學、金像獎 |
| TEXT07 | 心得、小說、興趣 | 小說、書 |
| TEXT08 | 實質、媒介、商業性、消費者、知識、訊息、特徵、出版物 | 出版物、價值、商業、廣告 |
| TEXT09 | 台北、專輯 | 稿、感情、點子、城市 |
| TEXT10 | 蘇聯、戈巴契夫、經濟、改革、補貼、物價、中小企業 | 蘇聯、經濟、改革、政策 |

**Table 4: Number of Subjects Selected by Readers**

|  | R1 | R2 | R3 | R4 | R5 | R6 | R7 | R8 | AVG | STDEV |
|---|---|---|---|---|---|---|---|---|---|---|
| TEXT01 | 3 | 4 | 7 | 2 | 3 | 5 | 4 | 7 | 4.38 | 1.85 |
| TEXT02 | 4 | 4 | 10 | 4 | 4 | 9 | 7 | 8 | 6.25 | 2.55 |
| TEXT03 | 4 | 5 | 4 | 10 | 4 | 8 | 8 | 7 | 6.25 | 2.31 |
| TEXT04 | 3 | 4 | 7 | 5 | 4 | 6 | 6 | 6 | 5.13 | 1.36 |
| TEXT05 | 2 | 4 | 7 | 6 | 4 | 8 | 6 | 7 | 5.50 | 2.00 |
| TEXT06 | 2 | 3 | 19 | 5 | 3 | 9 | 9 | 4 | 6.75 | 5.63 |
| TEXT07 | 2 | 4 | 8 | 3 | 2 | 6 | 6 | 4 | 4.38 | 2.13 |
| TEXT08 | 3 | 5 | 11 | 8 | 4 | 9 | 9 | 8 | 7.13 | 2.80 |
| TEXT09 | 3 | 3 | 8 | 2 | 4 | 5 | 5 | 4 | 4.25 | 1.83 |
| TEXT10 | 4 | 5 | 8 | 7 | 4 | 8 | 9 | 8 | 6.63 | 2.00 |
| AVG | 3.0 | 4.1 | 8.9 | 5.2 | 3.6 | 7.3 | 6.9 | 6.3 |  |  |
| STDEV | 0.82 | 0.74 | 4.01 | 2.62 | 0.70 | 1.64 | 1.79 | 1.70 |  |  |

**Table 5: The Repetition of Subjects Selected by Readers**

|  | 1 Reader | 2 Readers | 3 Readers | 4 Readers | 5 Readers | 6 Readers | 7 Readers | 8 Readers |
|---|---|---|---|---|---|---|---|---|
| TEXT01 | 2 | 4 | 1 | 0 | 0 | 1 | 0 | 2 |
| TEXT02 | 4 | 4 | 3 | 2 | 0 | 0 | 3 | 0 |
| TEXT03 | 4 | 7 | 0 | 1 | 2 | 3 | 0 | 0 |
| TEXT04 | 5 | 3 | 0 | 1 | 1 | 1 | 1 | 1 |
| TEXT05 | 0 | 3 | 1 | 1 | 2 | 2 | 0 | 1 |
| TEXT06 | 18 | 5 | 2 | 0 | 0 | 1 | 2 | 0 |
| TEXT07 | 5 | 1 | 4 | 0 | 2 | 1 | 0 | 0 |
| TEXT08 | 7 | 8 | 4 | 3 | 2 | 0 | 0 | 0 |
| TEXT09 | 4 | 3 | 4 | 0 | 1 | 0 | 1 | 0 |
| TEXT10 | 8 | 7 | 1 | 1 | 0 | 0 | 0 | 3 |
| SUM | 57 | 45 | 20 | 9 | 10 | 9 | 7 | 7 |

**Table 6: The Repetition of Subjects Selected by Readers and the Model**

|        | None | 1 Reader | 2 Readers | 3 Readers | 4 Readers | 5 Readers | 6 Readers | 7 Readers | 8 Readers |
|--------|------|----------|-----------|-----------|-----------|-----------|-----------|-----------|-----------|
| TEXT01 | 0    | 1        | 1         | 0         | 0         | 0         | 1         | 0         | 1         |
| TEXT02 | 1    | 1        | 2         | 1         | 2         | 0         | 0         | 1         | 0         |
| TEXT03 | 1    | 1        | 2         | 0         | 1         | 0         | 0         | 0         | 0         |
| TEXT04 | 5    | 1        | 1         | 0         | 1         | 0         | 0         | 1         | 1         |
| TEXT05 | 6    | 0        | 2         | 0         | 0         | 0         | 1         | 0         | 1         |
| TEXT06 | 5    | 3        | 0         | 0         | 0         | 0         | 0         | 0         | 0         |
| TEXT07 | 9    | 0        | 0         | 0         | 0         | 0         | 0         | 0         | 0         |
| TEXT08 | 2    | 2        | 2         | 0         | 2         | 0         | 0         | 0         | 0         |
| TEXT09 | 5    | 1        | 0         | 2         | 0         | 1         | 0         | 0         | 0         |
| TEXT10 | 4    | 3        | 1         | 0         | 0         | 0         | 0         | 0         | 1         |
| SUM    | 38   | 13       | 11        | 3         | 6         | 1         | 2         | 2         | 4         |

To compare the performance of the proposed model and that of readers, we show the number of the overlapping subjects selected by readers and identified by the model in Table 6. The second column of Table 6, "None", denotes that there is no overlapping subjects among readers and model. The third column, "1 Reader", denotes that one of the 8 readers is consistent with the model. Taking a look at Table 5 and Table 6 together, our model identifies the subjects selected by all readers in some texts, i.e., TEXT01, TEXT04, TEXT05, and TEXT10. However, the result of TEXT07 is not good. After examining the subjects selected by readers for TEXT07, we find that these subjects are very divergent. This result is in line with the aforementioned situation. To sum up, the distribution of data in Table 5 and that in Table 6 are consistent with each other. This shows that the performance of our model is comparable to that of readers. Moreover, it could reduce the costs taken in human efforts.

**CONCLUDING REMARKS**

This paper investigates the feasibility of automatic subject identification for textual documents. Through the proposed model and a series of experiments, the performance of our model shows the comparable power to that of human readers. Three factors of the reading and the writing of human beings, which are word frequency, word co-occurrence, and word distance, are introduced first. Another crucial factor, word importance, is domain-dependent. In general, the importance of a word is different in various types of texts. As a sequence, the measurement for word importance should be adaptive from domain to domain.

Basically, automatic subject identification would reduce the human efforts that are involved in libraries. However, from the viewpoint of library development (physical libraries or virtual libraries), it is necessary to provide numerous metadata for readers or users. The subject is only one possible metadata to describe DLOs. Considering the different information needs and the different metadata formats, it is important to introduce other automatic mechanisms for helping readers and users retrieve appropriate DLOs. For example, the date, time, place, and proper name are critical for DLOs and the automatic identification mechanisms for these informations will be very helpful. Researchers of information extraction have examined the impacts of these automatic mechanisms. Therefore, how to integrate the various language techniques becomes an important issue for the researches of information retrieval and information extraction.

**ACKNOWLEDGMENTS**
This work is supported in part by the National Science Council of R.O.C. under the contract NSC-86-2621-E002-025-T. I would like to thank CKIP Group, Academia Sinica for kindly providing Sinica Corpus.

**APPENDIX**

%% 文類=評論
%% 文體=論說
%% 語式=written
%% 主題=經濟
%% 媒體=報紙
%% 姓名=
%% 性別=男女
%% 國籍=中華民國
%% 母語=中文
%% 出版單位=中國時報
%% 出版地=臺灣
%% 出版日期=
%% 版次=
%% 日期=19910822

1. 。(PERIODCATEGORY) 藍騰銘(Nb) ８０年(Nd) ０８月(Nd) ２２日(Nd) 似乎(D) 緩不濟急(VH) 。(PERIODCATEGORY)

2. 。(PERIODCATEGORY) 最(Dfa) 大(VH) 冒險(VA)[+nom] 。(PERIODCATEGORY)

3. 。(PERIODCATEGORY) 因為(Cbb) 開放(VC) 改革(VC) 所(D) 帶來(VC) 的(DE) 不僅(Da) 是(SHI) 制度(Na) 上(Ng) 的(DE) 不同(VH)[+nom] ，(COMMACATEGORY)

4. ，(COMMACATEGORY) 這(Nep) 套(Nf) 制度(Na) 的(DE) 背後(Nc) 更(D) 蘊含(VJ) 著(Di) 一(Neu) 個(Nf) 不同(VH) 的(DE) 思想(Na) 體系(Na) 及(Caa) 行為(Na) 模式(Na) 。(PERIODCATEGORY)

5. 。(PERIODCATEGORY) 這些(Neqa) 思想(Na) 與(Caa) 行為(Na) ，(COMMACATEGORY)

6. ，(COMMACATEGORY) 與(P) 蘇聯(Nc) 舊有(A) 的(DE) 思考(VE)[+nom] 行為(Na) 模式(Na) 其實(D) 有(V_2) 著(Di) 許多(Neqa) 的(DE) 矛盾(VH)[+nom] ，(COMMACATEGORY)

7. ，(COMMACATEGORY) 而(Cbb) 這些(Neqa) 矛盾(VH)[+nom] 不但(Cbb) 反映(VE) 在(P) 蘇聯(Nc) 人民(Na) 無所適從(VI) 的(DE) 行為(Na) 上(Ng) ，(COMMACATEGORY)

8. ，(COMMACATEGORY) 同時(Nd) 也(D) 危及(VJ) 了(Di) 許多(Neqa) 特權(Na) 的(DE) 既得利益(Na) 。(PERIODCATEGORY)

9. 。(PERIODCATEGORY) 問題(Na) 重重(Neqa) 。(PERIODCATEGORY)

10. 。(PERIODCATEGORY) 但(Cbb) 經改(Na) 及(Caa) 開放(VC) 則(D) 帶領(VF) 蘇聯(Nc) 進入(VCL) 一(Neu) 個(Nf) 前所未有(VH) 的(DE) 轉型期(Na) ，(COMMACATEGORY)

11. ，(COMMACATEGORY) 而(Cbb) 轉型(VH) 時(Ng) 舊有(A) 體制(Na) 打破(VC) ，(COMMACATEGORY)

12. ，(COMMACATEGORY) 新(VH) 體制(Na) 建立(VC) 時(Ng) 勢必(D) 會(D) 危及(VJ) 某(Nes) 些(Nf) 特權(Na) 的(DE) 利益(Na) 。(PERIODCATEGORY)